\documentstyle[11pt,newpasp,twoside,epsf]{article} 

\begin{document} 

\title{Dark Matter in Dwarf Galaxies: High Resolution Observations} 
\author{Alberto D. Bolatto, Joshua D. Simon, Adam Leroy, \& Leo Blitz} 

\affil{Radio Astronomy Laboratory and Department of Astronomy,
University of California at Berkeley, 601 Campbell Hall, Berkeley, CA
94720, USA}

\begin{abstract}
We present observations and analysis of rotation curves and dark
matter halo density profiles in the central regions of four nearby
dwarf galaxies.  This observing program has been designed to overcome
some of the limitations of other rotation curve studies that rely
mostly on longslit spectra. We find that these objects exhibit the
full range of central density profiles between $\rho\propto r^0$
(constant density) and $\rho\propto r^{-1}$ (NFW halo). This result
suggests that there is a distribution of central density slopes rather
than a unique halo density profile.
\end{abstract}

\section{Introduction}

The last few years have seen a flurry of activity in the field of
precision measurements of central density profiles in dark matter
halos, as demonstrated elsewhere in these proceedings. Most of this
activity has concentrated on addressing two questions: is there a
unique central density profile slope?, and, more to the point, do the
measurements agree with the predictions of the simulations? In fact,
the observations appear to point to a substantial disagreement
between the central density profiles measured in low mass, low surface
brightness dwarf galaxies (e.g., de Blok et al. 2001) and the
predictions of most Cold Dark Matter simulations (e.g., Navarro,
Frenk, \& White 1996, hereafter NFW; Moore et al. 1999; Jing \& Suto
2000). The significance of this discrepancy, however, is a matter
of debate. Some authors ascribe it to intrinsic systematic
problems in the observations, which conspire to poorly constrain the
central density slopes (e.g., Swaters et al. 2003a), while others
acknowledge these problems but argue that the data provide strong
enough constraints to rule out universal slopes as steep as those
predicted by the simulations (e.g., de Blok, Bosma, \& McGaugh 2003).

Can observations be used to test the predictions of cosmological
simulations? In the current era of ``precision cosmology'' the answer
should be most emphatically {\em yes!} Close attention needs to be
paid to the potential systematic effects, however, in order to
minimize their importance. In this paper we present a series of
studies of high resolution velocity fields of dwarf galaxies which we
have designed to remove as much as possible the impact of systematic
uncertainties. Our overall conclusion is that (as also suggested by de
Blok et al. 2003) these objects appear to exhibit a range of central
density profiles rather than a unique value. The discrepancy between
observations and simulations is thus real, and perhaps related to the
absence of baryons and their related astrophysics in the simulations
(although other explanations are also possible; e.g., Ricotti
2003). In this regard, these observations should be taken as a reality
check on future simulations that incorporate all the physics relevant
on the small spatial scales.

\section{Experimental Design}

Several systematic problems have been identified in the literature as
potential causes of artificially shallow central density profiles. A
problem that plagued several earlier studies based on H{\small I}
observations was the lack of angular resolution and the consequent
smearing of the possible central density cusp. To avoid this pitfall,
and to attain the best possible angular and spatial resolution, our
program targets nearby dwarf galaxies with a combination of high
resolution millimeter interferometry (obtained at BIMA) and optical
spectroscopy. The use of two wavelengths allows us to avoid the
limitations inherent to one or another tracer: because CO emission is
faint and patchy, millimeter CO interferometry is signal--to--noise
limited and can only be used on a few objects. However, it does
provide 2D velocity fields and it is positionally extremely
accurate. Conversely, H$\alpha$ spectroscopy can be adversely affected
by obscuration and positioning problems.  In particular, incorrect or
inaccurate slit positioning in longslit spectra can cause
artificially flat rotation curves. To overcome this problem we acquire
integral field H$\alpha$ spectroscopy of our targets using the
DensePak multifiber spectrograph at the WIYN telescope.  To increase
the positional accuracy of the resulting H$\alpha$ velocity field we
cross--correlate the integrated intensity from the individual DensePak
footprints with a narrow--band H$\alpha$ image of the target.  The 2D
spectroscopic data allows us to study in detail the kinematics of the
galaxies, and in particular to look at the harmonic decomposition of the
velocity field in order to characterize their noncircular motions.
Finally, we use multiband optical and near--IR photometry to model and
remove the contribution of the stellar disk to the overall kinematics.

\begin{figure}[htb]
\plotone{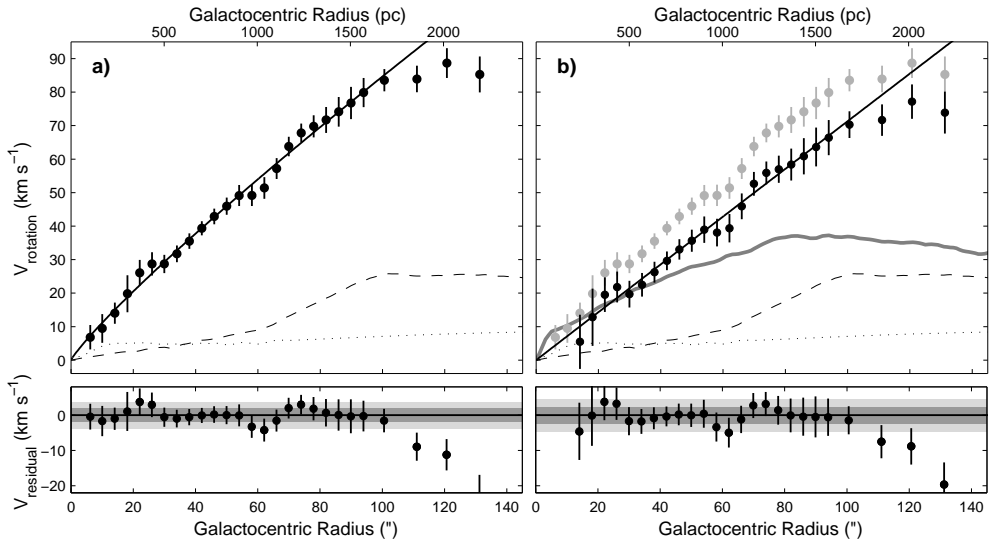}
\caption{Rotation curve of NGC~2976. Panel (a) shows the rotation
curve measured before removing any contribution from the disk
component (``minimum disk'' solution). The error bars incorporate the
uncertainties in the geometrical parameters used for the inversion, as
described in the main text. The dashed and dotted lines show the
circular velocities due to the H{\small I} and H$_2$ components of the
gaseous disk.  The lower panel shows the residuals after removing a
power law fit for $14\arcsec<r<105\arcsec$, with the gray regions
indicating $1\sigma$ and $2\sigma$ deviations ($\rho\propto r^{-0.27\pm0.09}$
assuming a spherical halo). Panel (b) shows the result of removing the
contributions of the gaseous and the stellar disk (thick gray line)
from the measured rotation. The stellar disk is maximal, with a
mass--to--light ratio $ M_*/L_K=0.19 M_\odot/L_{\odot K}$.  The fit to
the density profile shows it is constant density inside 1.8
kpc ($\rho\propto r^{-0.01\pm0.12}$). \label{n2976}}
\end{figure}

Once the velocity fields are obtained, the data analysis proceeds as
follows. The galaxy is deprojected and divided in concentric tilted
rings, with center, position angle, and axis ratio determined using
the available multiband photometry. A function of the form
$v_{sys}+v_{cir}\cos\theta+v_{rad}\sin\theta$ (where $\theta$ is the
deprojected position angle measured from the major axis) is fitted to
each ring, representing the effects of the systemic, circular, and
radial velocities. After making sure that the kinematics traced by the
radio and optical data agree, both datasets are combined to
obtain an overall velocity field. 

A crucial part of any rotation curve study is to quantify the errors,
which have direct bearing on the strenght of the constraints placed on
the shape of the density profile.  We use bootstrap simulations to
estimate the final errors in the center position, position angle, and
inclination of the galaxy; these errors then are obtained from the data
itself, and not derived from the fitted uncertainties of the individual
velocity measurements which can be unrealistically small.  
Finally, a Monte Carlo simulation with the
actual velocity field and these errors as input parameters is used to
determine the final uncertainties in the individual points of the
rotation curve. The rotation curve thus produced is the starting point
of the kinematic study.

The azimuthally averaged IR photometry of the galaxy is used to model
the potential of an infinitely thin stellar disk using Toomre's method.
%(e.g., Binney \& Tremaine 1987). 
The circular velocity due to this
disk is then subtracted in quadrature from the measured rotation
curve.  The mass--to--light ratio of the stellar disk can in principle
be obtained from the colors of the stellar population, either via
population synthesis models (such as the popular Starburst99), or
using empirical relationships (e.g., Bell \& de Jong 2001).  Similar
procedures are followed to estimate the contributions of the gaseous
disk to the rotation, when data are available. The rotation curve
obtained after removing the disk contributions is used to obtain the
dark matter density profile. The density profiles quoted here assume
spherical halos.

\begin{figure}[htb]
\plotone{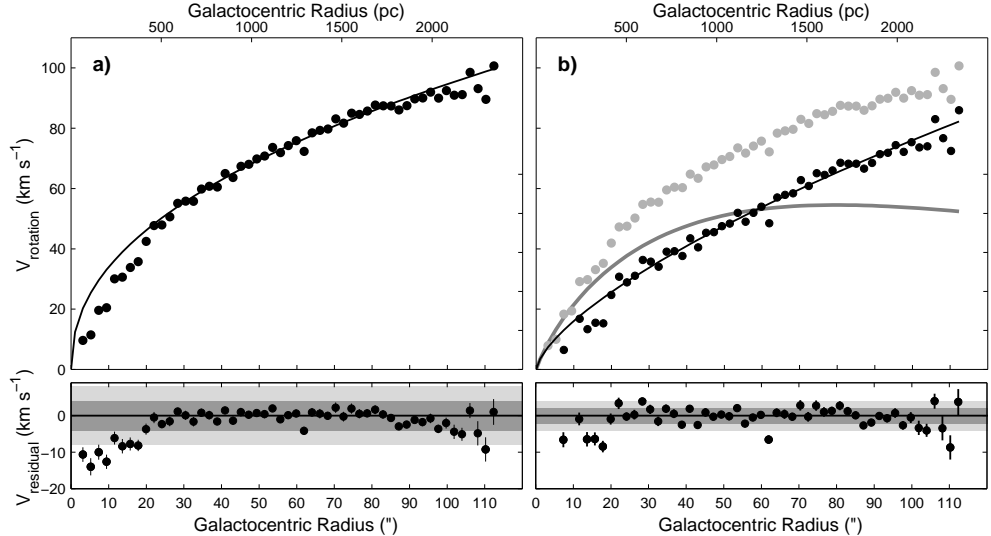}
\caption{Rotation curve of NGC~4605. Panel (a) shows the minimum
disk solution for this galaxy. The black line shows the power law
fit for $r>25\arcsec$ ($\rho\propto r^{-1.1}$), which significantly
overestimates the rotation velocities for the innermost points. These
are better fit with an almost constant density core ($\rho \propto
r^{-0.4}$). Panel (b) shows the maximum disk solution for this
galaxy. After subtracting the contribution of a maximal exponential
disk (thick gray curve), the density profile is fit by a $\rho\propto
r^{-0.65\pm0.1}$ power law at all radii. This study was carried out
using 2D CO velocity information and a 1D H$\alpha$ longslit
spectrum. We now have 2D H$\alpha$ and H{\small I} data, and plan to
revisit this object soon. \label{n4605}}
\end{figure}

\begin{figure}[htb]
\plotone{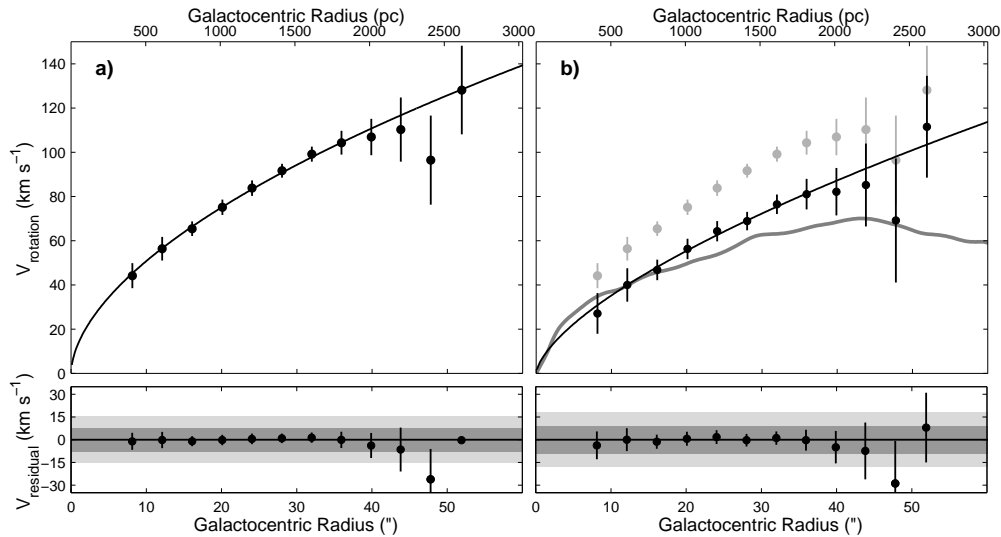}
\caption{Rotation curve of NGC~5949. Panel (a) shows the minimum disk
solution and its power law fit ($\rho\propto r^{-0.87\pm0.1}$).  Panel
(b) shows the result after removing the contribution from the disk
(thick gray curve). In this case we have used a submaximal disk with
$M_*/L_K=0.5 M_\odot/L_{\odot K}$, similar to what is observed in the
Milky Way (the maximal disk solution for NGC~5949 has
$M_*/L_K\simeq0.8 M_\odot/L_{\odot K}$). This is a compromise
solution: at the moment of writing these proceedings we lack enough
photometry information to better constrain the mass--to--light ratio
of this galaxy. With the chosen M/L, the power law fit to the dark
matter halo density profile is $\rho\propto
r^{-0.69\pm0.18}$.\label{n5949}}
\end{figure}

\section{Results}

\begin{figure}
\plotone{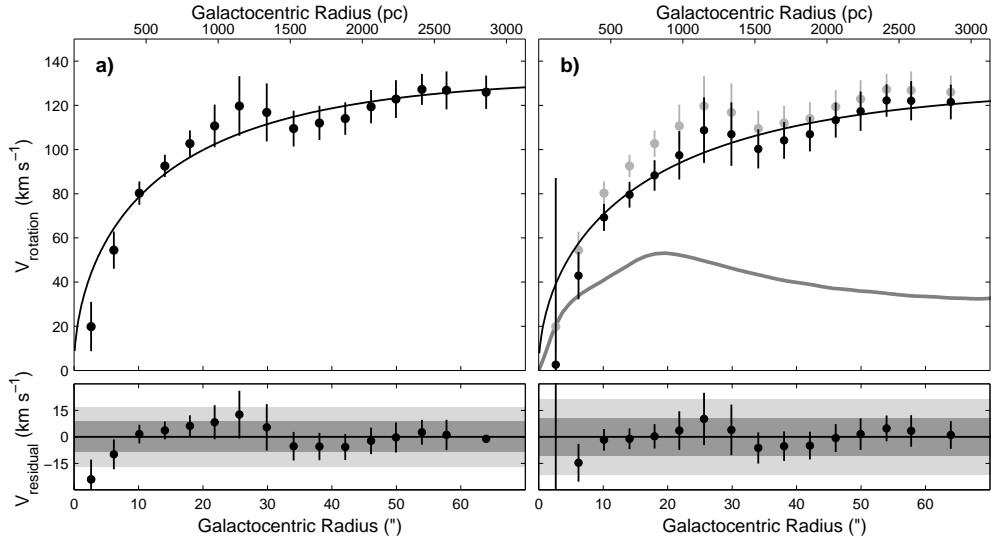}
\caption{Rotation curve of NGC~5963. Panel (a) shows the minimum disk
solution, together with an NFW fit to the measurements with a
concentration parameter of 20. This rotation curve is better fit by
the NFW functional form than by a power law. Panel (b) shows the
maximum disk solution, with $ M_*/L_I=0.7 M_\odot/L_{\odot I}$. The
resulting dark matter halo rotation curve is well--fit by the NFW
solution with R$_{\mbox{s}}=3.5$ kpc, V$_{200}=95$ km s$^{-1}$, and
R$_{200}=50$ kpc. \label{n5963}}
\end{figure}

We have completed the type of study described above in five nearby
dwarf galaxies: NGC~2976 (Simon et al. 2003; Fig. \ref{n2976}), NGC~4605
(Bolatto et al. 2002; Fig. \ref{n4605}), NGC~5949 (Fig. \ref{n5949}),
NGC~5963 (Fig. \ref{n5963}), and NGC~4625. The latter galaxy is part
of an interacting pair with the associated problems in interpreting
its velocity field (in hindsight, a poor choice).  In the other four
cases, however, we have been able to measure central density
profiles. Two of these galaxies (NGC~2976 and NGC~4605) have
measurable $v_{rad}$ terms that are usually associated with the
presence of bars. Analysis of the multicolor images, surface
photometry, and higher order harmonic decomposition of the kinematics
of NGC~2976, however, show no evidence for a bar. The radial motions
in NGC~4605 are comparatively less important, and it appears unlikely
that it hosts a bar (because of its higher inclination, however, this
case is less clear than that of NGC~2976). Neither NGC~5949 nor
NGC~5963 show measurable radial motions ($v_{rad}<5$ km s$^{-1}$),
although from the images the latter may contain a small bar.

\section{Conclusions}

The four galaxies in this series of studies for which we were able to
retrieve reliable central density profiles appear to span the full
range of behaviors; from constant density cores to NFW halos. In
particular, the data for NGC~2976 clearly do not allow an NFW halo.
Because we have tried to eliminate most of the systematics present in
this type of measurement, we believe that this range of central
density slopes is real and reflects an underlying distribution,
although the caveats associated with a small sample certainly
apply. To study this distribution we plan to expand the sample of
observed galaxies to 10--15 objects in the near future. This larger
sample will also allow us to look for correlations between central
density slopes and other parameters, such as galaxy mass or the
magnitude of noncircular motions.

Perhaps the most important conclusion, however, is that measurements
of the central density profiles of dwarf galaxies can be accurate.
Multiwavelength imaging spectroscopy is key to minimize the
vulnerability of these observations to potential systematic problems,
such as erroneous positioning of the spectrograph slit. At the same
time, 2D high--resolution data also provides a wealth of kinematic
information that would be otherwise unavailable.

Finally, the presence of measurable radial motions in two out of the
four galaxies studied presents a bit of a mystery; are we observing
the remnants of the processes that erased the original central cusps?
There does not appear to be a clear link to asymmetries or bars in our
small sample, and maintaining such motions over long time periods
appears very difficult in these small objects. Furthermore, it is
unclear what the effect of these motions is on the inversion the
rotation curve to obtain the density profile (we explicitly assumed
that they provide no ``support'', which is equivalent to ignoring
them). %Similar noncircular motions have been found in other small
%galaxies (e.g., Swaters et al. 2003b).  
Expanding the sample of
studied objects will allow us to determine how general this phenomenon
is.

 \end{document}